\documentclass[pra,twocolumn,titlepage,nofootinbib,amsmath,showpacs]{revtex4}%
\usepackage{amsmath}
\usepackage{amsfonts}
\usepackage{amssymb}
\usepackage{graphicx}%
\begin{document}
\title{\large Model-independent determination of the magnetic radius of
the proton from spectroscopy of ordinary and muonic hydrogen}
\author{Savely~G.~Karshenboim}
\email{savely.karshenboim@mpq.mpg.de}
\affiliation{Max-Planck-Institut f\"ur Quantenoptik, Garching,
85748, Germany} \affiliation{Pulkovo Observatory, St.Petersburg,
196140, Russia}

%\date

%\today

\begin{abstract}
To date the magnetic radius of the proton has been determined only by
means of electron-proton scattering, which is not free of controversies.
Any existing atomic determinations are irrelevant because they are
strongly model-dependent. We consider a so-called Zemach contribution
to the hyperfine interval in ordinary and muonic hydrogen and derive a
self-consistent model-independent value of the magnetic radius of
the proton. More accurately, we constrain not a value of the magnetic
radius by itself, but its certain combination with the
electric-charge radius of the proton, namely, $R_E^2+R_M^2$. The
result from the ordinary hydrogen is found to be
$R_E^2+R_M^2=1.35(12)\;{\rm fm}^2$, while the derived muonic value
is $1.49(18)\;{\rm fm}^2$. That allows us to constrain the value of
the magnetic radius of proton $R_M=0.78(8)\;{\rm fm}$ at the 10\% level.
\end{abstract}

%\today

%
\maketitle

\section{Introduction}

\begin{figure}[thbp]
\begin{center}
\resizebox{0.90\columnwidth}{!}{\includegraphics{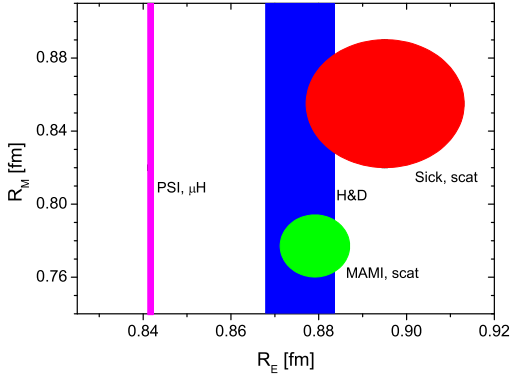} }
\end{center}
\caption{Determination of the rms proton charge and magnetic radii.
(Sick \cite{sick} has evaluated all the world data, but MAMI results
\cite{mami}. Other evaluations of those data produced similar
results (see, e.g., \cite{zhan}).) Ellipses for
electron-proton scattering should be somewhat turned from a pure
horizontal position because of a small correlation between $R_E$ and
$R_M$. For details see \cite{my_adp,my_ufn}.}
\label{fig:re}       % Give a unique label
\end{figure}

While a discrepancy between results on determination of the electric
charge radius of the proton has lately attracted attention of theoreticians
and experimentalists, a controversy in
determination of the magnetic radius is rather in shadow. The
situation is summarized in Fig.~\ref{fig:re}. The
proton charge radius has already been discussed in \cite{efit}, which
is referred here as paper I. This paper is a direct continuation of paper I
and we do not reproduce here any plots
or equations from there.

A stronger interest to the situation with the electric charge radius $R_E$
is due to a broader variety of the data and more important
applications, such as determination of the Rydberg constant. While
in the case of the magnetic radius $R_M$ there is a discrepancy
between two scattering results \cite{mami,sick}, the set of results
for $R_E$ also includes  the spectroscopic data on hydrogen and
deuterium \cite{codata2010} and on muonic hydrogen
\cite{Nature,Science}.

The contradiction between different values of
$R_M$ is rather serious, while reading the published results literally.
To certain extent it is expected that the discrepancy is partly due to
different treatment of the proton polarizability
contribution~\cite{comm,resp}. However, that may remove only a part of
the discrepancy.

The spectroscopic data and, in particular, results on the
hyperfine-structure (HFS) interval in hydrogen ($1s$) and muonic hydrogen
($2s$) \cite{Science}, may present a source for an independent extraction
of $R_M$, however, at present there is no model-independent constraints
on $R_M$ from the HFS interval in muonic or ordinary hydrogen.
Values published from time to time are deduced from models of the proton
form factors, but there has been no realistic model of the proton
developed to date.

Any comparison of `pure' QED theory with the experimental results on
hydrogen has been `contaminated' for decades by the presence of certain proton
finite-size and polarizability contributions. While the experimental value
of the $1s$ interval in hydrogen had been for a while among the most
accurately measured physical quantities, the most sensitive QED
tests for the HFS-interval theory has been performed not
with the $1s$ interval in hydrogen, but with quantities free of the
influence of the nuclear structure. Such quantities have been provided
by a study of leptonic atoms, such as muonium or positronium. Another
opportunity is a comparison of the $1s$ and $2s$ HFS intervals
measured with the same atom. Details of those QED tests with the HFS
can be found, e.g., in review \cite{my_rep}.

Here we explore a related question. The main purpose of this note is
to estimate the constraints on the magnetic radius of the proton, $R_M$,
from the hyperfine splitting in muonic and ordinary hydrogen. If the proton
polarizability contributions are known with a sufficient accuracy, we
can experimentally determine the value of the proton finite-size
contribution by a comparison of the theory and experiment. Such a
contribution must be sensitive to the distribution of both electric charge
and magnetic moment inside the proton.
Considering that contribution in an appropriate way, we intend to extract a
constraint on a certain combination of $R_E$ and $R_M$.

While the QED effects are well understood (see, e.g.,
\cite{my_rep}), the total theoretical accuracy for the HFS interval in
both muonic and ordinary hydrogen is
completely determined by the proton-structure terms, namely, by the
elastic two-photon contribution and by the proton polarizability
correction. In case of hydrogen the experimental uncertainty is
negligible, while for $\mu$H it is compatible with and somewhat
higher than the theoretical one.

As for calculation of the elastic term, its dominant part can be found
in the external field approximation. We have to deal with integral
\begin{eqnarray}\label{i1def}
I_1^{\rm EM}&\equiv&  \int_0^\infty
{\frac{dq}{q^2}}\left[\frac{G_E(q^2)G_M(q^2)}{\mu_p}-1\right]\;,\label{def:i1}
\end{eqnarray}
which determines the dominant proton-finite-size contributions into the HFS
interval in ordinary and muonic hydrogen
\begin{eqnarray}
\Delta E_{\rm HFS}(ns) &=&\frac{8(Z\alpha)m_r}{\pi n^3} E_F
\,I_1^{\rm EM}
%\nonumber\\
% &=&-2(Z\alpha)m E_F\,\langle r \rangle_Z
\;,
\end{eqnarray}
where $E_F$ is the so-called Fermi energy, $m_r$ is the reduced mass
of a bound electron (in hydrogen) or muon (in muonic hydrogen) and
$\mu_p=2.7928...$ is the proton magnetic moment in units of the
nuclear magnetons. For available experimental data $n=1$ for
ordinary hydrogen\footnote{The $2s$ HFS interval in hydrogen is also
well measured \cite{2s}. The experimental accuracy is worse than for the
$1s$, however, it still supersedes the theoretical accuracy. The $1s$
and $2s$ data are consistent and a separate consideration of the
$2s$ HFS interval would not add any new information on the proton
structure. Meanwhile, comparison of the $1s$ and $2s$ results allows a
sensitive test of QED  (see, e.g., \cite{my_rep}).} (see, e.g., a
summary on the $1s$ HFS interval in \cite{1s}) and $n=2$ for muonic
hydrogen \cite{Science}. The other notation used for the integral
under question presents it in terms of the so-called {\em Zemach radius\/}
(or {\em the first Zemach momentum\/})
\begin{eqnarray}
\langle r \rangle_Z &=&-\frac{4}{\pi}I_1^{\rm EM}\;.
\end{eqnarray}

We have no direct experimental knowledge on the integrand in
(\ref{i1def}), which consists of the subtracted form factors of the proton,
${G_E(q^2)G_M(q^2)}/{\mu_p}-1$. In particular, the accurate data
fail at low momenta, which essentially contribute to the integral.
Everything used in the integrand was a result of certain fitting rather than
direct  measurements. (We in part explore here ideas
presented previously in \cite{pla} and developed in paper I.)

\begin{figure}[thbp]
\begin{center}
\resizebox{0.8\columnwidth}{!}{\includegraphics{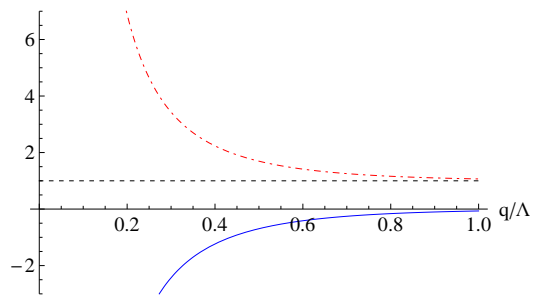}}
\end{center}
\caption{Fractional contributions to the integrand in (\ref{def:i1})
as a function of $q/\Lambda$ as follows from the dipole model. The red
dot-dashed line is the subtraction term with unity and the blue solid line
is the $G_EG_M$ term, i.e. for the data (cf. \cite{pla,efit}).}
\label{fig:ind1}       % Give a unique label
\end{figure}

The situation with the integrand in (\ref{i1def}) is illustrated in
Fig.~\ref{fig:ind1}, where various fractional contributions to the
integrand are estimated from the dipole model and presented as a
function of $q/\Lambda$. The red dot-dashed line is for the subtraction term
with unity. The blue solid line is for the $G_EG_M$ term, which is related to
the data. The integral is fast convergent at high $q$. At low
$q$, say below $0.3\Lambda$, the data contribution produces a large
uncertainty and any successful result for the Zemach contribution
obtained previously was based on a certain, sometimes unrealistic, model.

We are going to split the integration into two parts:
\begin{equation}
I=\int_0^\infty {{dq}{...}}\equiv I_<+I_>\equiv
\int_0^{q_0}{{dq}{...}} + \int_{q_0}^\infty {{dq}{...}}
\end{equation}
which are to be treated differently (cf. \cite{pla,efit}).

For higher momenta, we will use direct experimental data (or rather their
realistic approximation). The
accuracy of the form factors is roughly 1\%. The integral over the
direct data is indeed singular at $q_0\to0$, because the
experimental values of $G_E(0)$ and $G_M(0)/\mu_p$ are not equal to
unity exactly --- they are only consistent with unity within the
uncertainty, which produces the singularity. The smaller is $q_0$
the larger is the uncertainty of the related integral.

On the other hand, we can expand the form factors at low momentum
\begin{equation}\label{g2}
\frac{G_E(q^2)G_M(q^2)}{\mu_p}= A + B q^2 + C q^4 + ...\nonumber\\
\end{equation}
Some contributions into $I_{1<}^{\rm EM}$ vanish because of the subtraction and
the uncertainty comes from the remaining terms. The smaller is $q_0$
the smaller is the uncertainty. Here, $A=1$ and
$B=-(R_E^2+R_M^2)/6$. The $B$ contribution to $I_{1<}^{\rm EM}$
\begin{equation}
I_1^{\rm
R}=-\frac{R_E^2+R_M^2}{6}q_0
\end{equation}
is to be treated separately. That is the `signal' that we use to
constrain $R_E^2+R_M^2$. The
leading remaining term is the $C$ term, which is responsible for
the uncertainty.

The idea is to apply a certain model to estimate the uncertainties
and to find a value of $q_0$, which corresponds to the smallest
uncertainty possible (cf. paper I).

Concluding on the model to estimate the uncertainty, we note that
the dipole form factor is a reasonable estimation for the form
factors as far as we discuss general features, but not any accurate
particular value. So, we can, e.g., set for $G_EG_M/\mu_p$
\begin{eqnarray}\label{def:b}
C&=&bC^{\rm dip}\;,\nonumber\\
C^{\rm dip}&=&\frac{10}{\Lambda^4}\;,
\end{eqnarray}
and estimate the $b$ coefficient as $b=1\pm1$ (cf. \cite{efit}).
Here, we use for various preliminary estimations the standard dipole model
\[
G_{\rm dip}(q^2)=\left(\frac{\Lambda^2}{q^2+\Lambda^2}\right)^2
\]
and apply for numerical evaluations $\Lambda^2=0.71\,{\rm GeV}^2$,
which corresponds to $R_{\rm dip}=0.811\;$fm.

\section{Consideration within the dipole model}

Let us perform an evaluation of $I_1^{\rm EM}$ following the consideration of
$I_3^{\rm E}$ in paper I.

The complete dipole value useful for further estimation of the
fractional uncertainties is
\begin{eqnarray}
I_1^{\rm dip}&=&  \int_0^\infty {\frac{dq}{q^2}}\left[\left(G_d(q^2)\right)^2-1\right]\nonumber\\
&=&-\frac{35}{32} \frac\pi\Lambda\nonumber\\
&\simeq&-4.08\;{\rm GeV}^{-1}\nonumber\\
&\simeq& -0.805\;{\rm fm}\;.\label{eq:dip}
\end{eqnarray}

\section{Splitting the integral into parts}

As we intend to split the integral into two parts, let us start with
the higher-momentum part
\begin{eqnarray}
I_{1>}^{\rm EM}&=&  \int_{q_0}^\infty {\frac{dq}{q^2}}\left[\frac{G_E(q^2)G_M(q^2)}{\mu_p}-1\right]\nonumber\\
&=& \int_{q_0}^\infty {\frac{dq}{q^2}\frac{G_E(q^2)G_M(q^2)}{\mu_p}}
- \frac{1}{q_0}\;.
\end{eqnarray}
Its uncertainty is estimated, by considering the part of the integral,
singular at the limit $q_0\to0$. The result is
\begin{eqnarray}\label{eq:dgg}
\delta I_{1>}^{\rm EM}&=& \delta \int_{q_0}^\infty {\frac{dq}{q^2}}\frac{G_E(q^2)G_M(q^2)}{\mu_p}\nonumber\\
&\simeq&  \delta \int_{q_0}^\infty {\frac{dq}{q^2}}\frac{G_E(q_0^2)G_M(q_0^2)}{\mu_p}\nonumber\\
&=& \frac{1}{\nu \Lambda}\frac{2\delta G(q_0^2)}{G(q_0^2)}
\left(\frac{1}{1+\nu^2}\right)^4
\end{eqnarray}
or
\[
\frac{\delta I_{1>}^{\rm EM}}{-I_1^{\rm dip}}\simeq
\frac{0.0058}{\nu}\left(\frac{1}{1+\nu^2}\right)^4\;,
\]
where $\nu=q_0/\Lambda$ and we suggest for our estimations that
both electric and magnetic form factors roughly follow the
standard dipole fit and we experimentally know both of them within
1\% uncertainty
\[
\frac{\delta G(q_0^2)}{G(q_0^2)}\equiv\frac{\delta
G_E(q_0^2)}{G_E(q_0^2)}\simeq\frac{\delta
G_M(q_0^2)}{G_M(q_0^2)}\simeq1\%\;.
\]
Here we apply the dipole values for estimation of absolute and
fractional uncertainties (cf. paper I).

Meanwhile, at low momenta, we find
\begin{eqnarray}
I_{1<}^{\rm EM}&=&  \int_0^{q_0} {\frac{dq}{q^2}}\left[\frac{G_E(q^2)G_M(q^2)}{\mu_p}-1\right]\nonumber\\
&=& -\frac{R_E^2+R_M^2}{6}q_0 + \frac{10}{3} b
\frac{q_0^3}{\Lambda^4}\;,
\end{eqnarray}
where $b$ was defined above.

\section{The extraction: a general consideration}

Combining an experimental value, QED contributions and a polarizability
correction we obtain
\begin{eqnarray}
I_1^{\rm exp}&=&\frac{\pi n^3}{8(Z\alpha)m_rE_F} \left( E_{\rm
HFS}^{\rm exp}-E_{\rm HFS}^{\rm QED}-\Delta E_{\rm HFS}^{\rm
polarizability}
\right)\nonumber\\
&=&\frac{\pi}{8(Z\alpha)m_r} \frac{\left( E_{\rm HFS}^{\rm
exp}-E_{\rm HFS}^{\rm QED}-\Delta E_{\rm HFS}^{\rm polarizability}
\right)}{E_{\rm HFS}}
\end{eqnarray}
where we noted that $E_{\rm HFS}\simeq E_F/n^3$ with accuracy sufficient
for the denominator. Alternatively, we can write
\begin{equation}
\langle r \rangle_Z^{\rm exp} =-\frac{1}{2(Z\alpha)m_r} \frac{\left(
E_{\rm HFS}^{\rm exp}-E_{\rm HFS}^{\rm QED}-\Delta E_{\rm HFS}^{\rm
polarizability} \right)}{E_{\rm HFS}}\;.
\end{equation}
Indeed, there are also some higher-order proton-structure
corrections, such as a recoil part of the two-photon exchange. We
assume that they are included if necessary in the QED or
polarizability term.

Often in some papers, they present $\langle r \rangle_Z^{\rm
exp}$ rather than $I_1^{\rm exp}$. Some `experimental' values of $I_1^{\rm
exp}$ are summarized in Table~\ref{t:i1exp}. Any `experimental'
value is a result of an extraction procedure that deeply involves
theory and, in particular, a calculation of the proton polarizability,
which dominates in the uncertainty budget for hydrogen and produces an
uncertainty comparable with the measurement uncertainty for muonic
hydrogen.

\begin{table}[htbp]
\begin{center}
\begin{tabular}{l|c|c|c|c|c}
\hline
Atom & State & $\langle r \rangle_Z$ & $I_1^{\rm exp}$ & $\delta I_1^{\rm exp}/I_1^{\rm exp}$ & Ref. \\[0.8ex]
\hline
H      & $1s$ & 1.047(16) fm &  $-4.17(6)\,{\rm GeV}^{-1}$   &1.5\%  &  \cite{h30} \\[0.8ex]
H      & $1s$ & 1.037(16) fm & $-4.13(6)\,{\rm GeV}^{-1}$   & 1.5\% & \cite{h31} \\[0.8ex]
$\mu$H & $2s$ & 1.082(37) fm  &  $-4.31(15)\,{\rm GeV}^{-1}$   &3.4\% & \cite{Science} \\[0.8ex]
\hline
\end{tabular}
\caption{Some `experimental' values for $I_1$. For numerical evaluations
in this paper we use for hydrogen the value from  \cite{h30}.
\label{t:i1exp}}
 \end{center}
 \end{table}

Meantime, according to our theoretical consideration
\begin{eqnarray}
I_1^{\rm th}&=&-\frac{R_E^2+R_M^2}{6}q_0 + \frac{10}{3}
\bigl(1\pm1\bigr)
\frac{q_0^3}{\Lambda^4}+I_{1>}^{\rm EM}\nonumber\\
&=&-\frac{R_E^2+R_M^2}{6}q_0 +\left(I_1^{\rm EM}-I_1^{\rm R}\right)\;,
\end{eqnarray}
and thus we arrive at
\[
R_E^2+R_M^2=-\frac{6}{q_0}\Bigl(I_1^{\rm exp}-\left(I_1^{\rm EM}-I_1^{\rm
R}\right)\Bigr)\;.
\]

\section{The extraction: the uncertainty budget and its optimization}

It may be useful to introduce the fractional uncertainty of
$R_E^2+R_M^2$
\[
\delta_{2R}=\frac{\delta \bigl( R_E^2+R_M^2\bigr)}{R_E^2+R_M^2}\;.
\]
Since roughly $R_E\approx R_M \approx R_{\rm dip}$, a somewhat
different value
\begin{eqnarray}
\delta'_{2R}&=&\frac{\delta \bigl( R_E^2+R_M^2\bigr)}{2R_{\rm
dip}^2}
\end{eqnarray}
is roughly equal to $\delta_{2R}$, but easier to handle. It is sufficient to minimize the uncertainty of determination of $R_E^2+R_M^2$.

It is equal to the rms sum of partial uncertainties, which are
\begin{eqnarray}
\delta'_{\rm exp}&=&\frac{6}{q_0}\frac{I_1^{\rm dip}}{2R_{\rm
dip}^2}\frac{\delta I_1^{\rm exp} }{I_1^{\rm dip}}\;,\nonumber\\
\delta'_{<}&=&\frac{6}{q_0}\frac{I_1^{\rm dip}}{2R_{\rm dip}^2}
\frac{\delta I_{1<}^{\rm EM} }{I_1^{\rm dip}}\;,
\nonumber\\
\delta'_{>}&=&\frac{6}{q_0}\frac{I_1^{\rm dip}}{2R_{\rm dip}^2}
\frac{\delta I_{1>}^{\rm EM}}{I_1^{\rm dip}}\nonumber\;.
\end{eqnarray}

With
\[
R_{\rm dip}^2=\frac{12}{\Lambda^2}
\]
and
\[
\frac{6}{\Lambda}\frac{-I_1^{\rm dip}}{2R_{\rm dip}^2}=0.8590...\;,
\]
we obtain
\begin{eqnarray}
\delta'_{<}&=&0.83\,\nu^2
\nonumber\;,\\
\delta'_{>}&=&\frac{0.0050}{
\nu^2}\left(\frac{1}{1+\nu^2}\right)^4\nonumber\;.
\end{eqnarray}

To find $\delta_{\rm exp}$, we have to utilize the results from
Table~\ref{t:i1exp}
\begin{eqnarray}
\delta'_{\rm exp,\; H}&=&\frac{0.013}{\nu}\;,\nonumber\\
\delta'_{{\rm exp,\;}\mu{\rm H}}&=&\frac{0.031}{\nu}\nonumber\;,
\end{eqnarray}
where for ordinary hydrogen we use the result from \cite{h30}.

%\section{extraction: the best value of $\nu_0$}

\begin{figure}[thbp]
\begin{center}
\resizebox{0.80\columnwidth}{!}{\includegraphics{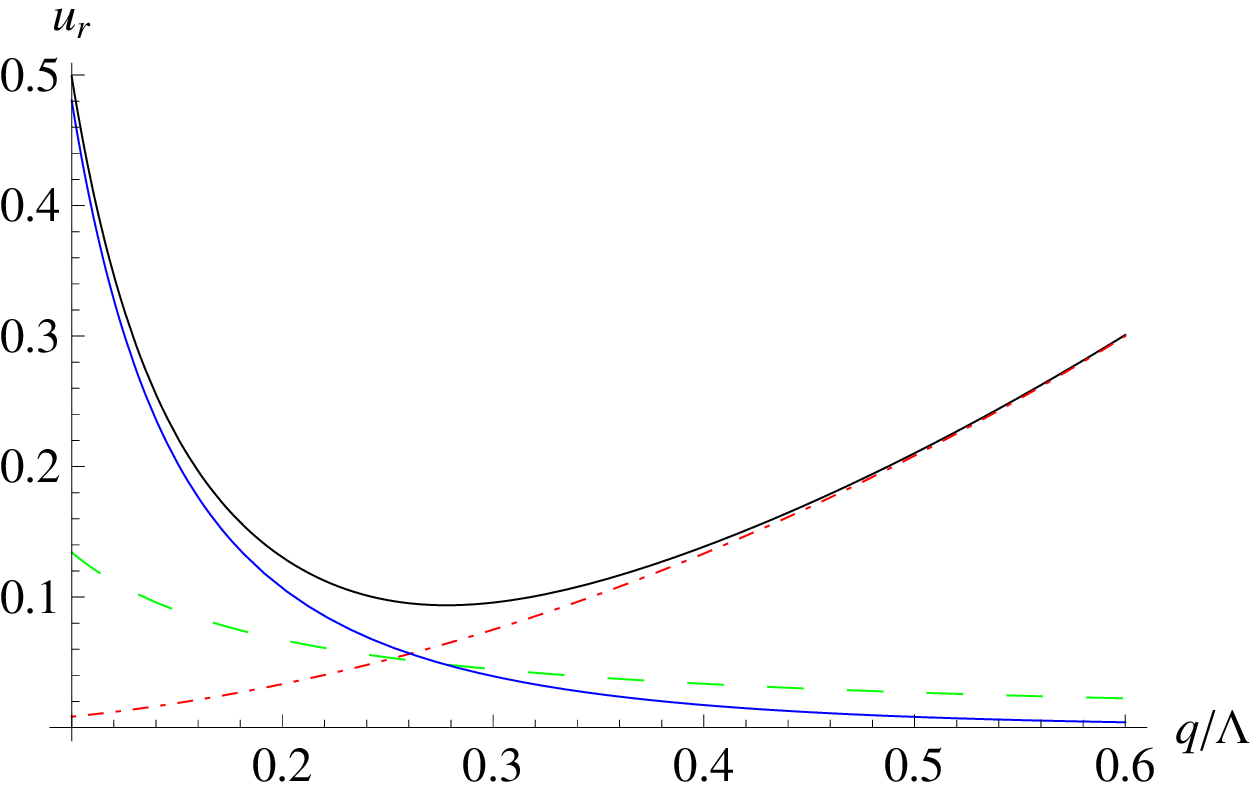}}
\resizebox{0.80\columnwidth}{!}{\includegraphics{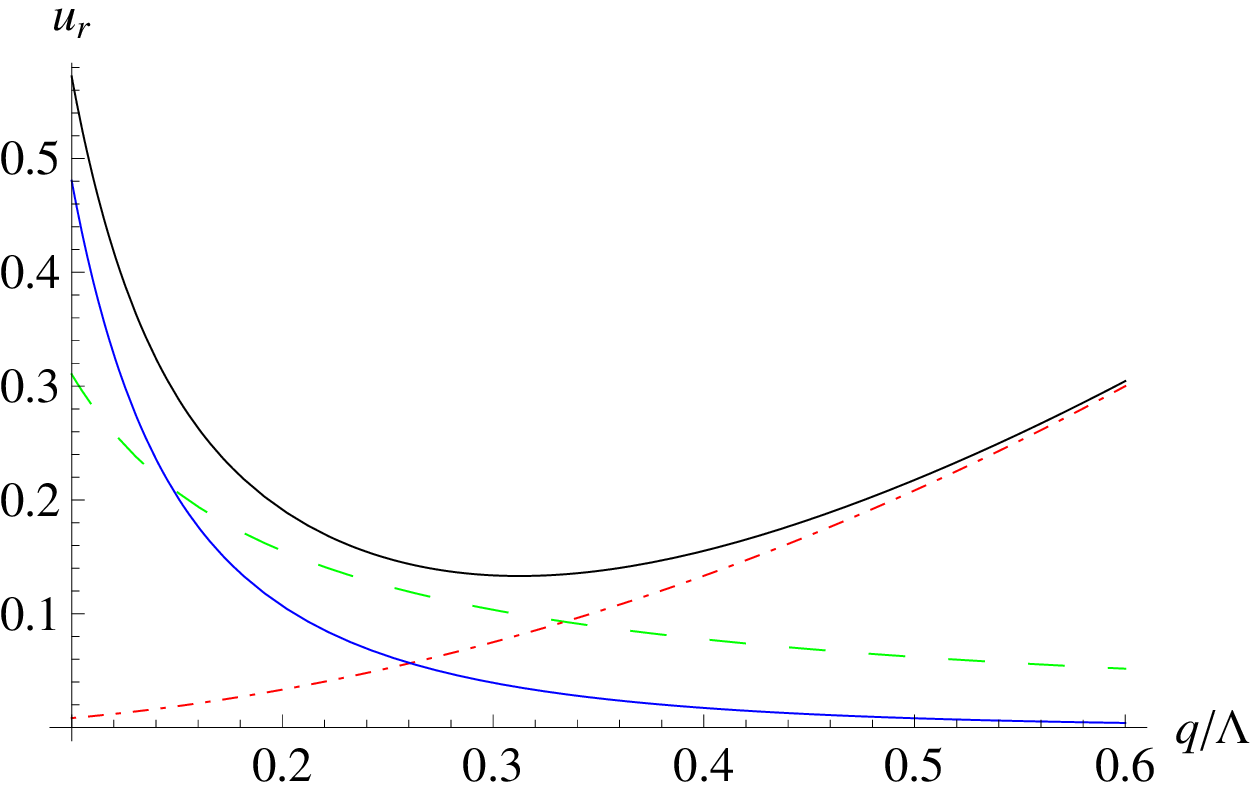}}
\end{center}
\caption{The total fractional uncertainty $\delta'_{2R}$ (a black solid line)
as the rms
sum of sources listed above as a function of $\nu=q_0/\Lambda$. The
partial uncertainties are also presented. The top plot is for
ordinary hydrogen, while the bottom one is for $\mu$H. The red dot-dashed lines
are for the uncertainty of the low-momentum part of the $I_1^{\rm EM}$
integral and the blue solid ones are for high-momentum contribution to the
uncertainty budget; the `experimental' uncertainties, which are
different for muonic and ordinary hydrogen (see
Table~\ref{t:i1exp}), are presented with green dashed lines.}
\label{fig:unc1H}       % Give a unique label
\end{figure}

With the uncertainty determined, let us consider behavior of the
uncertainty as a function of $\nu=q_0/\Lambda$. All the partial
uncertainties as well as the total one as a function of $\nu$ are
plotted in Fig.~\ref{fig:unc1H} both for ordinary (top) and muonic
(bottom) hydrogen.

The optimal values, which minimize the uncertainty, and the partial
contributions to the total uncertainty for those values are
collected in Table~\ref{t:i1budget}.

\begin{table}[htbp]
\begin{center}
\begin{tabular}{l|c|c|c|c|c|c|c|c}
\hline Atom & Best $\nu$ & Best $q_0$ & $\delta'_{2R}$ &
$\delta'_{\rm exp}$ & $\delta'_{<}$ & $\delta'_{>}$  & {\em Scat}&
{\em Scat}$^*$
\\[0.8ex]
\hline
H      & 0.278 & 0.234 GeV &  9.4\%&  4.8\%&  6.4\%&  4.8\% &{\em 1.3\%}&{\em 3.2\%}\\[0.8ex]
$\mu$H & 0.312 & 0.263 GeV &  13.3\%&  9.9\%&  8.1\%&  3.5\% &{\em 0.9\%}&{\em 2.0\%}\\[0.8ex]
\hline
\end{tabular}
\caption{Parameters for evaluation of the fractional of
uncertainties and fractional scatter of the results (see below). The scatter
of the results from a fit to fit is used to control accuracy. It is not
included into the error budget. {\em Scatter\/} is for the scatter of the
results without fits from \cite{bo1994} and {\em scatter\/}$^*$ is for the
scatter including the fits  from \cite{bo1994} (see below for detail).
\label{t:i1budget}}
 \end{center}
 \end{table}

\section{The fits of the proton form factors}

Now we are to find $I_1^{\rm EM}-I_1^{\rm R}$ by integration over the data.
As in paper I, we use for that a certain set of fits.

As an approximation we utilize the fits for the proton form factors
$G_E$ and $G_M$ from
Arrington and Sick, 2007 \cite{as2007}, Kelly, 2004 \cite{kelly},
Arrington et al., 2007 \cite{am2007}, Alberico et al., 2009
\cite{ab2009}, Venkat et al., 2011 \cite{va2011}, and from
Bosted, 1995 \cite{bo1994}. The details of the fits for the electric
form factor are presented in \cite{efit}, while for the magnetic one
they are summarized in Appendix~\ref{s:fit}.

Two of the fits for $G_M$ are with so-called chain fractions, five
are with Pad\'e approximations with polynomials in $q^2$, and one is
a Pad\'e approximation with polynomials in $q$.

As well as in case of a pure electric integral in \cite{efit}, the
fit (\ref{fit:mbo1994}) of Bosted \cite{bo1994} is perfect for the
tests. It is a Pad\'e approximation with polynomials in $q$, not in
$q^2$. It definitely has a low-momentum behavior strongly different
from others.

\begin{figure}[thbp]
\begin{center}
\resizebox{0.85\columnwidth}{!}{\includegraphics{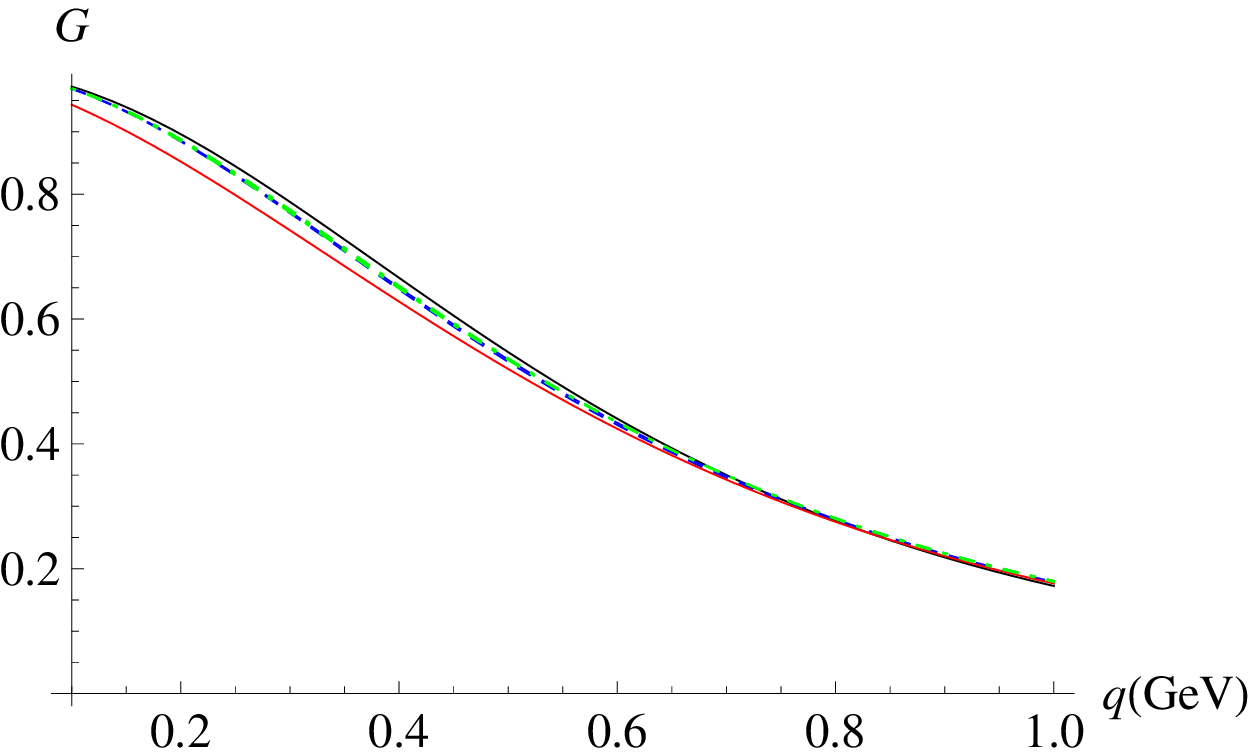}}
\resizebox{0.85\columnwidth}{!}{\includegraphics{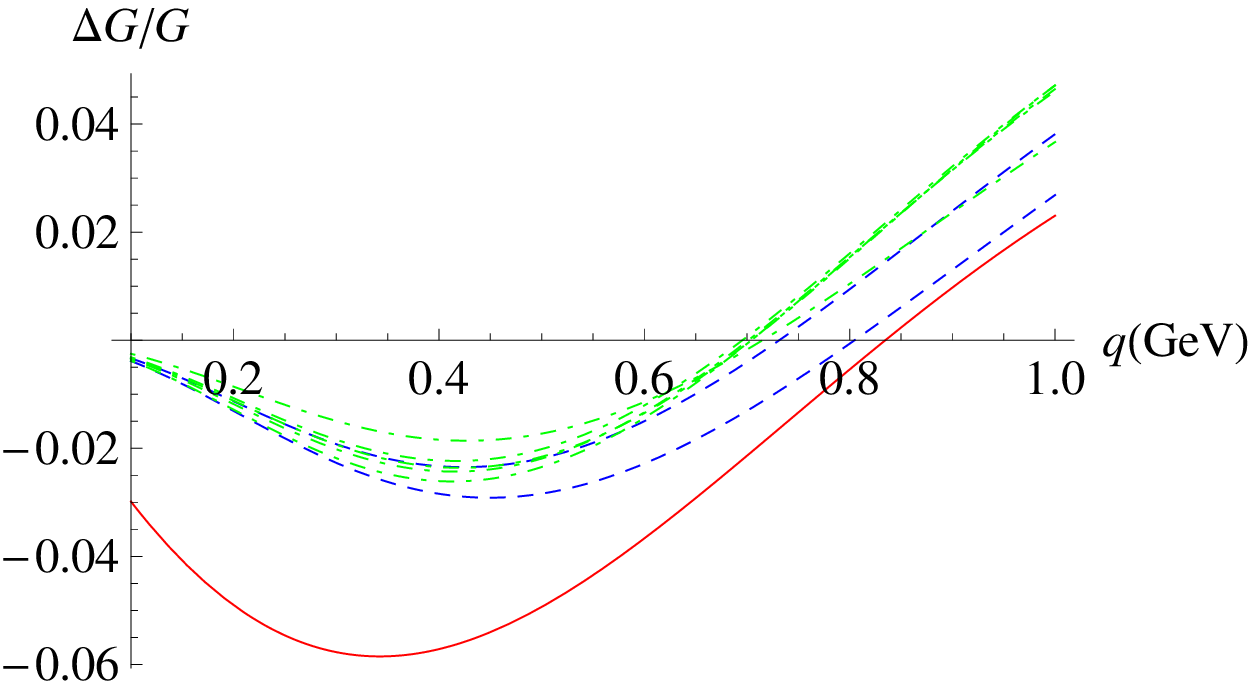} }
\end{center}
\caption{Top: Magnetic form factor $G_M(q^2)/\mu_p$ of the proton:
dipole parametrization and the fits. Bottom: Fractional deviation of
fits from the dipole form factor  $(G_M/\mu_p-G_{\rm dip})/G_{\rm
dip}$. Horizontal axis: $q\;$[Gev/$c$]. The blue dashed lines are for the chain
fractions, the green dot-dashed lines are for Pad\'e approximations with
$\tau=q^2/4m_p^2$ and the red solid one is for the Pad\'e approximation
with $q$. The dipole fit in the top graph is presented with a black solid line.}
\label{fig:mdip}       % Give a unique label
\end{figure}

\begin{figure}[thbp]
\begin{center}
\resizebox{0.85\columnwidth}{!}{\includegraphics{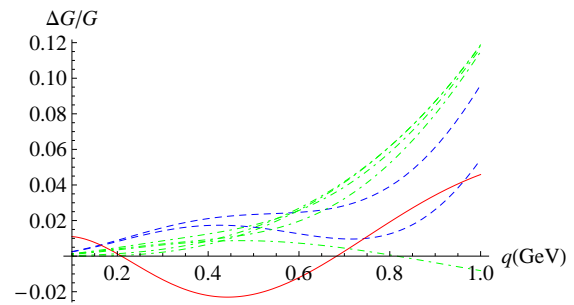}}
\end{center}
\caption{Fractional deviation of the magnetic form factor from the
electric one $(G_E-G_M/\mu_p)/G_E$ (from the same fitting
procedures). Horizontal axis: $q\;$[Gev/$c$]. The blue dashed lines are for the
chain fractions, the green dot-dashed lines are for Pad\'e approximations with
$\tau=q^2/4m_p^2$ and the red solid one is for the Pad\'e approximation
with $q$.}
\label{fig:e-m}       % Give a unique label
\end{figure}

The fits are quite close to one another and to the standard dipole
parametrization in the area of interest. They are more or less
consistent to each other and to the dipole one (see
Fig.~\ref{fig:mdip}). Comparison of the fits for the magnetic
form factor to the related fit for the electric form factor is also
presented (see, Fig.~\ref{fig:e-m}).

The low-momentum behavior of the fits is summarized in
Table~\ref{t:exp:mfits}. Indeed, the fit (\ref{fit:mbo1994}) from
\cite{bo1994} is excluded.

\begin{table}[htbp]
\begin{center}
\begin{tabular}{l|c|c|c|c|c}
\hline
Ref. & ~Fit & Type &  $R_E$ &  $R_M$  &  ~$C$ \\[0.8ex]
&&&[fm]& [fm]&[GeV$^{-4}$]\\
 \hline
\cite{as2007}&(\ref{fit:mas2007}) &  Chain fraction & 0.90 &0.86& 31.2 \\[0.8ex]
\cite{as2007}& (\ref{fit:mbm2005}) & Chain fraction & 0.90 & 0.87 & 32.2 \\[0.8ex]
\cite{kelly}& (\ref{fit:mkelly}) & Pad\'e approximation ($q^2$) & 0.86  &0.85 & 26.8 \\[0.8ex]
\cite{am2007}&(\ref{fit:mam2007}) &  Pad\'e approximation ($q^2$) & 0.88  &0.86& 29.2 \\[0.8ex]
\cite{ab2009}&(\ref{fit:mab20091}) &  Pad\'e approximation ($q^2$) & 0.87 &0.87& 28.7 \\[0.8ex]
\cite{ab2009}&(\ref{fit:mab20092}) &  Pad\'e approximation ($q^2$) & 0.87 &0.86& 28.2 \\[0.8ex]
\cite{va2011}&(\ref{fit:mva2011}) &  Pad\'e approximation ($q^2$) & 0.88 &0.86& 29.5 \\[0.8ex]
\hline
\end{tabular}
\caption{The low-momentum expansion of the fits can be expressed as
$G_E G_M/\mu_p= 1 - (R_E^2+R_M^2) \,q^2/3 + C q^4 +
...$. The values in the Table are given for central values of the
fits without any uncertainty. The references are given to the papers
where both electric and magnetic form factors are presented. The
summary on the applied fits for the electric form factor of the
proton can be found in \cite{efit}. The references to the
equations are given here for the magnetic form factor (see Appendix
\ref{s:fit}). The related values for the standard dipole fit are
$R_E=R_M=0.811\;{\rm fm}$ and $C=19.8\;{\rm
GeV}^{-4}$.\label{t:exp:mfits}}
 \end{center}
 \end{table}

\section{Integration over the fits}

Integrating over the fits for the optimal $q_0({\rm H})= 0.222\;$GeV
for hydrogen, we find that $I_{1>}^{\rm EM}$ varies from $-2.92\,$GeV$^{-1}$ to
$-2.89\,$GeV$^{-1}$ if we exclude Pad\'e approximation in $q$
\cite{bo1994} or from $-2.97\,$GeV$^{-1}$ if we include it. For
detail see Table~\ref{t:i1:fit}.
% \footnote{Here, we use the following
% notation for the scatter in the tables. Just `scatter' is for the
% scatter without applying (\ref{fit:mbo1994}) from \cite{bo1994} and
% `scatter$^*$ stands for the scatter with applying that fit.}.
Here,
we accept
\[
I_{1>}^{\rm EM}({\rm H})=-2.90(2)\;{\rm GeV}^{-1}
\]
as the mean value (excluding (\ref{fit:mbo1994})), that leads to
\[
I_1^{\rm EM}({\rm H})-I_1^R({\rm H})=-2.82(11)\;{\rm GeV}^{-1}\;.
\]
The uncertainty of integral above does not include scattering in the
calculation of $I_{1>}^{\rm EM}$ because we estimated the uncertainty of
this term in a more conservative way as explained above.

\begin{table}[htbp]
\begin{center}
\begin{tabular}{l|c|c}
\hline
Fit & Type & $I_{1>}^{\rm EM}$  \\[0.8ex]
\hline
\cite{as2007}& Chain fraction & $-2.91\;{\rm GeV}^{-1}$\\[0.8ex]
\cite{as2007}& Chain fraction & $-2.92\;{\rm GeV}^{-1}$\\[0.8ex]%former bm2005
\cite{kelly} & Pad\'e approximation ($q^2$) & $-2.89\;{\rm GeV}^{-1}$\\[0.8ex]
\cite{am2007}& Pad\'e approximation ($q^2$) & $-2.90\;{\rm GeV}^{-1}$\\[0.8ex]
\cite{ab2009}& Pad\'e approximation ($q^2$) & $-2.90\;{\rm GeV}^{-1}$\\[0.8ex]
\cite{ab2009}& Pad\'e approximation ($q^2$) & $-2.90\;{\rm GeV}^{-1}$\\[0.8ex]
\cite{va2011}& Pad\'e approximation ($q^2$) & $-2.90\;{\rm GeV}^{-1}$\\[0.8ex]
\cite{bo1994}& Pad\'e approximation ($q$) & $-2.97\;{\rm GeV}^{-1}$\\[0.8ex]
\hline
\end{tabular}
\caption{Scatter of data for  $I_{1>}^{\rm EM}$ for hydrogen at
`optimal' $q_0\simeq 0.278\,\Lambda=0.234\,{\rm
GeV}/c$.\label{t:i1:fit}}
 \end{center}
 \end{table}

Eventually, we obtain a constraint on the magnetic radius from the HFS
interval in hydrogen
\begin{equation}
\frac{R_E^2+R_M^2}{2R_{\rm dip}^2} = 1.025(94) \;,
\end{equation}
and we remind that for the standard dipole parametrization $R_{\rm
dip}^2=0.658\;{\rm fm}^2$.

The related fractional scatter is 0.013 if we exclude (\ref{fit:mbo1994}) from
the consideration and it is 0.032 if we include it. The result
for the combination of the proton electric and magnetic radius is
consistent with the value from the standard dipole model within its
9\% uncertainty.

The same evaluation can be performed for various $q_0$ and the
results are summarized in Table~\ref{t:un1h:q}. All the results are
consistent. The scatter is below the uncertainty except for very low
$q_0$, where behavior of the fits becomes model dependent.

\begin{table}[htbp]
\begin{center}
\begin{tabular}{l|c|c|c|c}
\hline
 $q_0/\Lambda$ &$q_0$ & $({R_E^2+R_M^2})/{2R_{\rm dip}^2}$ &{\em Scatter} &
{\em Scatter}$^*$ \\[0.8ex]
\hline
 0.20 & 0.169\;GeV & 0.99(13) & {\em 0.03} & {\em 0.10} \\[0.8ex]
 0.25 & 0.211\;GeV & 1.02(9) & {\em 0.02}& {\em 0.05} \\[0.8ex]
 0.30 & 0.253\;GeV & 1.03(10) & {\em 0.01}& {\em 0.02} \\[0.8ex]
 0.35 & 0.295\;GeV & 1.04(11) & {\em 0.007}& {\em 0.01} \\[0.8ex]
 0.40 & 0.337\;GeV & 1.05(14) & {\em 0.004}& {\em 0.007} \\[0.8ex]
 0.50 & 0.421\;GeV & 1.08(21) & {\em 0.002}& {\em 0.002} \\[0.8ex]
\hline
\end{tabular}
\caption{The results for $({R_E^2+R_M^2})/{2R_{\rm dip}^2}$ at
various $q_0$ for hydrogen. \label{t:un1h:q}}
 \end{center}
 \end{table}

Similar treatment for muonic hydrogen produces $q_0(\mu{\rm H})=
0.263\;$GeV as the optimized value. The results of integration over
the fits for $I_{1>}^{\rm EM}$ vary from $-2.77\:$GeV$^{-1}$ excluding
(\ref{fit:mbo1994}) and  from $-2.80\:$GeV$^{-1}$ including it to
$-2.74\:$GeV$^{-1}$ (see Table~\ref{t:i1:fit:mu}). We consider
\[
I_{1>}^{\rm EM}(\mu{\rm H})=-2.75(1)\;{\rm GeV}^{-1}
\]
as the mean value that leads to
\[
I_1^{\rm EM}(\mu{\rm H})-I_1^R(\mu{\rm H})=-2.63(13)\;{\rm
GeV}^{-1}\;.
\]
The uncertainty of integral above does not include scattering in a
calculation of $I_>$ because we estimated the uncertainty of this term
in a more conservative way as explained above.

\begin{table}[htbp]
\begin{center}
\begin{tabular}{l|c|c}
\hline
Fit & Type &   $I_{1>}^{\rm EM}$ \\[0.8ex]
\hline
\cite{as2007}& Chain fraction & $-2.76\;{\rm GeV}^{-3}$\\[0.8ex]
\cite{as2007}& Chain fraction & $-2.77\;{\rm GeV}^{-3}$\\[0.8ex]%former bm2005
\cite{kelly}& Pad\'e approximation ($q^2$) & $-2.74\;{\rm GeV}^{-3}$\\[0.8ex]
\cite{am2007}& Pad\'e approximation ($q^2$) & $-2.75\;{\rm GeV}^{-3}$\\[0.8ex]
\cite{ab2009}& Pad\'e approximation ($q^2$) & $-2.76\;{\rm GeV}^{-3}$\\[0.8ex]
\cite{ab2009}& Pad\'e approximation ($q^2$) & $-2.75\;{\rm GeV}^{-3}$\\[0.8ex]
\cite{va2011}& Pad\'e approximation ($q^2$) & $-2.75\;{\rm GeV}^{-3}$\\[0.8ex]
\cite{bo1994}& Pad\'e approximation ($q$) & $-2.80\;{\rm GeV}^{-3}$\\[0.8ex]
\hline
\end{tabular}
\caption{Scatter of data for  $I_{1>}^{\rm EM}$ for muonic hydrogen
at `optimal' $q_0\simeq 0.312\Lambda=0.263\,{\rm
GeV}/c$.\label{t:i1:fit:mu}}
 \end{center}
 \end{table}

The constraint from the HFS interval in muonic hydrogen is found to
be
\begin{equation}
\frac{R_E^2+R_M^2}{2R_{\rm dip}^2}= 1.13(13) \;.
%1.130(133)
\end{equation}
The fractional scatter is 0.009 (excluding (\ref{fit:mbo1994})) or 0.020
(including (\ref{fit:mbo1994})). The result is consistent with the
value from the standard dipole moment within the uncertainty of
13\%.

The results obtained at various values of the separation parameter
$q_0$ are consistent to each other (see Table~\ref{t:un1muh:q} for details).

 \begin{table}[htbp]
\begin{center}
\begin{tabular}{l|c|c|c|c}
\hline
 $q_0/\Lambda$ &$q_0$ & $({R_E^2+R_M^2})/{2R_{\rm dip}^2}$ &{\em scatter} &{\em scatter}$^*$ \\[0.8ex]
\hline
 0.20 & 0.169\;GeV & 1.14(19) & {\em 0.03} & {\em 0.10} \\[0.8ex]
 0.25 & 0.211\;GeV & 1.13(14) & {\em 0.02}& {\em 0.05} \\[0.8ex]
 0.30 & 0.253\;GeV & 1.13(13) & {\em 0.01}& {\em 0.02} \\[0.8ex]
 0.35 & 0.295\;GeV & 1.13(14) & {\em 0.007}& {\em 0.01} \\[0.8ex]
 0.40 & 0.337\;GeV & 1.13(16) & {\em 0.004}& {\em 0.007} \\[0.8ex]
 0.50 & 0.421\;GeV & 1.14(22) & {\em 0.002}& {\em 0.002} \\[0.8ex]
\hline
\end{tabular}
\caption{The results for $({R_E^2+R_M^2})/{2R_{\rm dip}^2}$ at various
$q_0$ for muonic hydrogen. \label{t:un1muh:q}}
 \end{center}
 \end{table}

\section{Conclusions}

Our strategy to evaluate $I_1^{\rm EM}$ was dictated by our purpose,
which is to determine $R_M$ (cf. \cite{pla}). For a different purpose
the strategy would be different.

\begin{figure}[thbp]
\begin{center}
\resizebox{0.99\columnwidth}{!}{\includegraphics{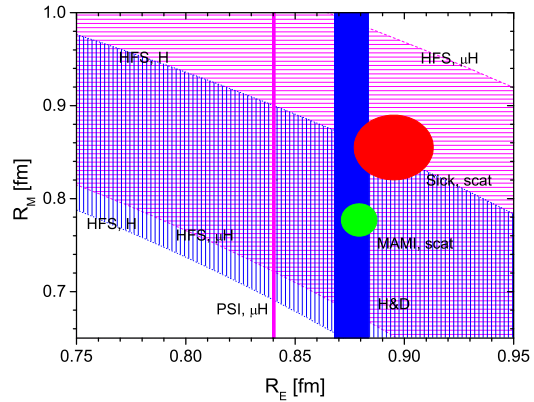} }
\end{center}
\caption{Determination of the rms proton charge and magnetic radii.
Notation is the same as in Fig.~\ref{fig:re}. Two new striped belts are
added from the HFS constraints in ordinary and muonic hydrogen
above. The former is filled with vertical lines and the latter is with
the horizontal ones. Their colors are the same as for the Lamb shift
in the related
atomic system. The squared area is the overlap of two striped areas.}
\label{fig:re:new}       % Give a unique label
\end{figure}

To obtain
a constraint on the magnetic radius of the proton $R_M$, the
compilation of all constraints on the electromagnetic radii of the proton
is presented in
Fig.~\ref{fig:re:new}. We plot there constraints from
Fig.~\ref{fig:re} and also present two constraints on $R_E^2+R_M^2$
derived here from a study of the HFS intervals in ordinary and muonic
hydrogen.

That is a general picture. It already has certain features similar
to those considered in paper I. The overall accuracy of
spectroscopic extractions of the magnetic radius looks comparable
with scattering
results---not with their claimed uncertainty, but with their
discrepancy.

The preliminary results on the proton magnetic radius from atomic
spectroscopy are presented in Table~\ref{t:rm}. It
involves all possible combinations of spectroscopic constraints on
$R_E^2+R_M^2$ (from HFS) and on $R_E$ (from the Lamb shift). Note
that ${\delta R_M}/{R_M}\approx\delta^\prime_{2R}$, assuming that
$R_E$ is known with a good accuracy and that roughly $R_E\approx R_M
\approx R_{\rm dip}$.

 \begin{table}[htbp]
 \begin{center}
\begin{tabular}{l|l|l}
\hline
Transition/atom            & Lamb ($\mu$H)          & Lamb (H)          \\[0.8ex]
 \hline
HFS (H)             &    0.80(8)  fm    & 0.76(8) fm        \\[0.8ex]
HFS ($\mu$H)            & 0.88(11) fm         & 0.85(11) fm        \\[0.8ex]
\hline
 \end{tabular}
\caption{Magnetic radius of the proton from combining HFS and the
Lamb shift in muonic and ordinary hydrogen.\label{t:rm}}
 \end{center}
 \end{table}

If we accept the value of the proton charge radius as
\[
R_E=0.86(2)\;{\rm fm}\;,
\]
which seems a reasonable choice until the controversy in its determination
is not resolved, then we arrive at
\[
R_M=0.78(8)\;{\rm fm}\;,
\]
as to the best constraint on the magnetic radius from spectroscopy.

\begin{figure}[thbp]
\begin{center}
\resizebox{0.85\columnwidth}{!}{\includegraphics{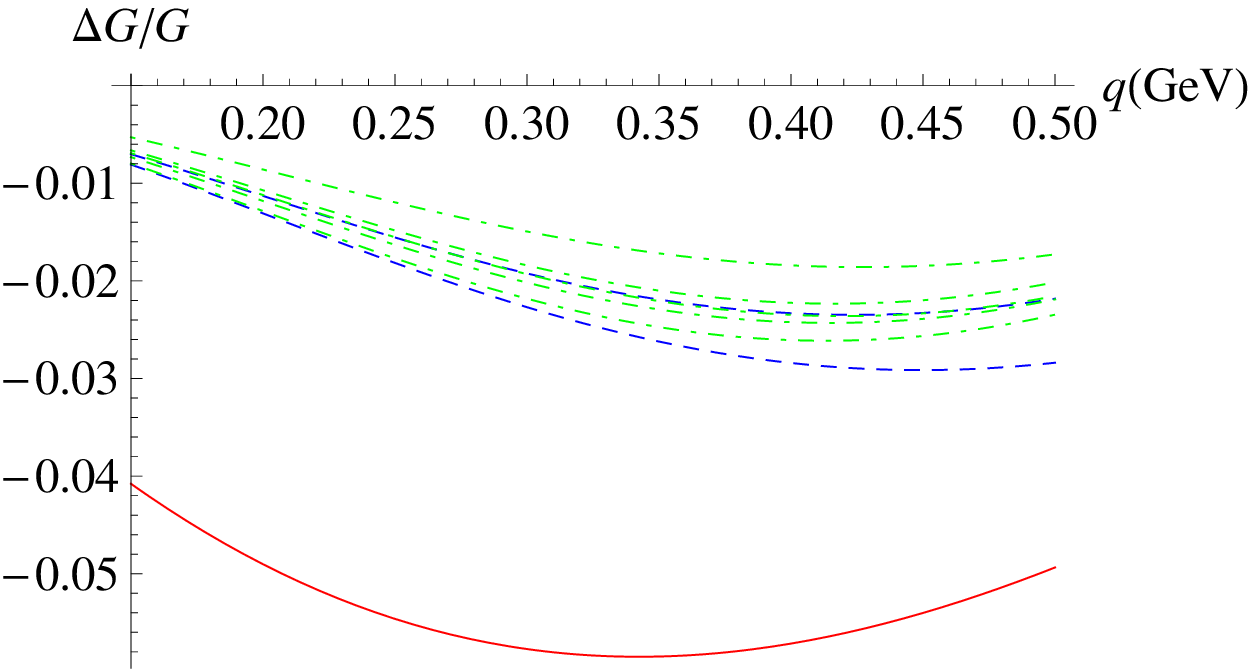} }
\resizebox{0.85\columnwidth}{!}{\includegraphics{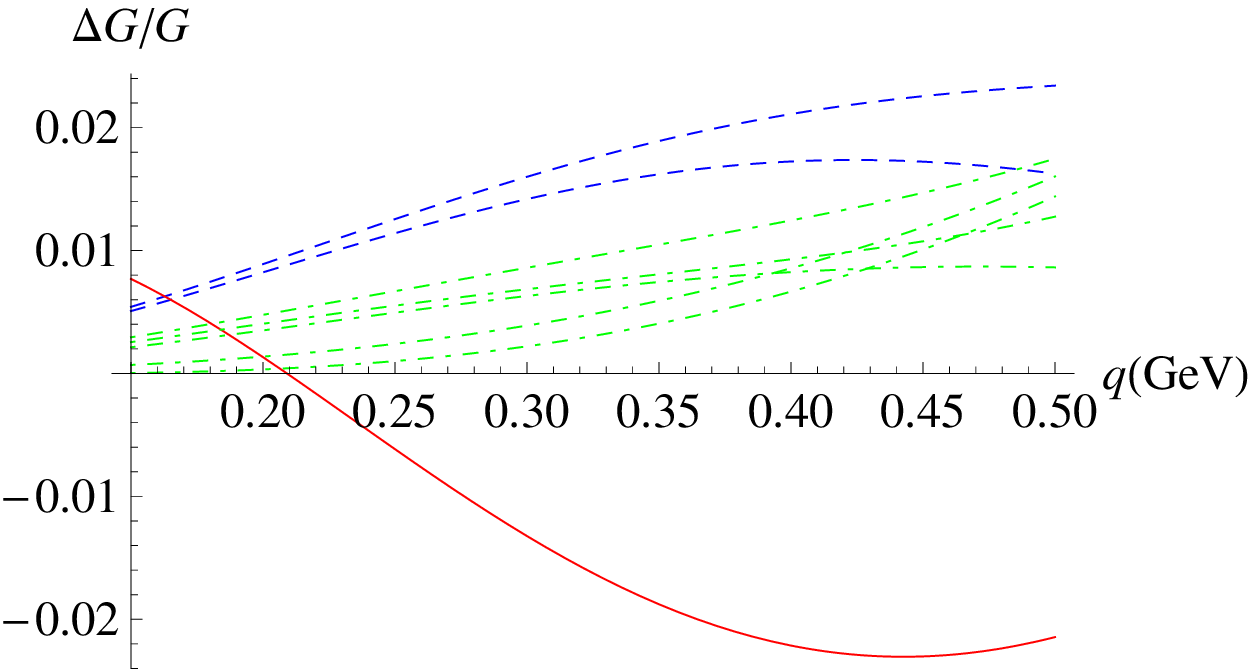} }
\end{center}
\caption{Top: Fractional deviation of the magnetic form factor from the
dipole form factor $(G_M/\mu_p-G_{\rm dip})/G_{\rm dip}$. Bottom:
Fractional deviation of the magnetic form factor from the electric
one $(G_E-G_M(q^2)/\mu_p)/G_E$. Horizontal axis: $q\;$[Gev/$c$] and
the range $0.2-0.5\;{\rm GeV}/c$ is crucial for $q_0$ in all three
cases considered.}
\label{fig:40}       % Give a unique label
\end{figure}

We have performed a number of consistency checks described above such as
consideration of various values of the separation parameter. All the
results are consistent. The estimation of the $q^4$ term is consistent with
all fits with reasonable behavior at low $q$ discussed in Appendix
(as well as with the fits from MAMI (see \cite{thesis} for detail)).
In case of the fits considered above the only fit with unreasonable behavior
is that from  \cite{bo1994}, which in particular produces infinite values
of the charge and mangetic radii. It is important that all fits but
(\ref{fit:mbo1994}) agree with each other within at 1\% level in the region
where the separation parameter was chosen in Tables~\ref{t:un1h:q} and
\ref{t:un1muh:q} as seen in Fig.~\ref{fig:40}.

The fit of \cite{bo1994} with incorrect behavior at low momentum
transfer is responsible for a scatter of values of $G_M(q_0)$ bigger
than $\pm1\%$ (see Fig.~\ref{fig:40}) (cf. \cite{efit}). However,
taking into account its unrealistic behavior, that is acceptable. The
other fits agree with each other within 1\% in the region crucial for a
choice of $q_0$. A similar situation with the electric radius (see
Fig.~6 of paper I).

A value of the magnetic radius extracted from electron-proton scattering
strongly depends on
treatment of the proton polarizability \cite{comm,resp}. To apply
the Rosenbluth separation, one has to rely on a certain model for the
proton polarizability. We note, however, that the electric form factor
$G_E$ at low momentum transfer is less
sensitive to the model and as far as we are not going to go too low,
one may use experimental results on $G_M/G_E$ from recoil polarimetry
(see, e.g.,
\cite{zhan}), which are free of the polarizability problem.

Similarly to  the case of examination for $R_E$ in \cite{efit}, we
conclude that the estimation of the uncertainty of the form factors
at the level of 1\% for applied $q_0$ is validated by the behavior
of the fits and by the scale of the scatter. Nevertheless, a direct
investigation of the problem would be useful.

The author is grateful to S. Eidelman and V. Ivanov for useful discussions.
This work was supported in part by DFG under grant HA 1457/9-1.

\appendix

\section{Fits for $G_M$ applied in the paper\label{app:fit}\label{s:fit}}

The fits for $G_M$ applied in the papers fall into three classes.

1). Two fits deal with chain fractions. Those are from from
Arrington and Sick, 2007, \cite{as2007}. One is from analysis
of \cite{as2007} alone\footnote{Here, $Q$ is the numerical
value for momentum transfer $q$ in GeV.}
\begin{eqnarray}\label{fit:mas2007}
\frac{G_M(Q^2)}{\mu_p}&=& \frac{1}{1 + \frac{3.173
        Q^2}{1 - \frac{0.314 Q^2}{ 1 - \frac{1.165Q^2}{1 + 5.619\frac{Q^2}{1 -
        1.087Q^2}}}}}\;,
\end{eqnarray}
while the other exploit the evaluation of the two-photon effects from
% Blunden %and Melnitchouk et al., 2005,
\cite{bm2005}
\begin{eqnarray}\label{fit:mbm2005}
\frac{G_M(q^2)}{\mu_p}&=& \frac{1}{1 + \frac{3.224
        Q^2}{1 - \frac{0.313 Q^2 }{ 1 - \frac{0.868 Q^2}{1 + \frac{4.278Q^2}{1 -
        1.102Q^2}}}}}\;,
\end{eqnarray}
and

Five fits are Pad\'e approximations with polynomials in $q^2$. Those
include fits from Kelly, 2004, \cite{kelly}
\begin{eqnarray}\label{fit:mkelly}
\frac{G_M(q^2)}{\mu_p}&=&\frac{1 + 0.12 \tau }{1 + 10.97 \tau +
18.86 \tau^2 +
      6.55 \tau^3}\;,
\end{eqnarray}
where
\[
\tau = q^2/4m_p^2\;,
\]
from Arrington %and Melnitchouk
et al., 2007, \cite{am2007}
\begin{eqnarray}\label{fit:mam2007}
\frac{G_M(q^2)}{\mu_p}&=& \frac{1 - 1.465 \tau + 1.260 \tau^2 +
0.262 \tau^3}{1 + 9.627 \tau + 11.179 \tau^4 +
      13.245 \tau ^5}\;,
\end{eqnarray}
from Alberico %and Bilenky
et al., 2009, \cite{ab2009}
\begin{eqnarray}\label{fit:mab20091}
\frac{G_M(q^2)}{\mu_p}&=& \frac{1 + 1.53\tau}{1 + 12.87 \tau +
29.16\tau^2 + 41.40
\tau^3}\\
\frac{G_M(q^2)}{\mu_p}&=&\frac{1 + 1.09\tau}{1 + 12.31 \tau +
25.57\tau^2 + 30.61 \tau^3}\label{fit:mab20092}\;,
\end{eqnarray}
and two fits Venkat %and Arrington
et al., 2011, \cite{va2011}
\begin{eqnarray}\label{fit:mva2011}
\frac{G_M(q^2)}{\mu_p}&=&\frac{1-1.43573\tau+1.19052066\tau^2+0.25455841\tau^3}{D_{V}}\;,\nonumber\\
D_V&=&1+9.70703681\tau+3.7357\times10^{-4}\tau^2\nonumber\\
&&+6.0\times10^{-8}\tau^3+ 9.9527277\tau^4\nonumber\\
&&+12.7977739\tau^5\;.
\end{eqnarray}

The remaining fit from Bosted, 1995, \cite{bo1994}
\begin{eqnarray}\label{fit:mbo1994}
\frac{G_M(q^2)}{\mu_p}&=& \left(1 + 0.35Q + 2.44 Q^2 + 0.50 Q^3\right.\nonumber\\
&&+\left.
1.04Q^4 + 0.34Q^5\right)^{-1}
\end{eqnarray}
is a Pad\'e approximation with polynomials in $q$. That is a
phenomenological fit designed to be used for medium and high $q$. It
is not expected to be appropriate at low $q$. Providing a reasonably
good approximation at medium momentum transfer, the fit apparently
has incorrect low-$q$ behavior and incorrect analytic properties
such as a branch point at $q^2=0$.

\end{document}